\documentstyle[prl,aps,amsfonts]{revtex}

\title{Disordered Dirac Fermions: Multifractality Termination and
  Logarithmic Conformal Field Theories}

\author {J.-S. Caux$^{1}$, N. Taniguchi$^{1,2}$, and A. M. Tsvelik$^{1}$}

\address{$^1$ Department of Physics, University of Oxford, 1 Keble Road
  Oxford,OX1 3NP, UK\\ $^2$ Department of Physical Electronics, Hiroshima
  University, Higashi-Hiroshima 739, Japan}

\date{7 January 1998}

\begin{document}
\draft
\maketitle

\begin{abstract}
  We reexamine in detail the problem of fermions interacting with a
  non-Abelian random vector potential.  Without resorting to the replica
  or supersymmetry approaches, we show that in the limit of infinite
  disorder strength the theory possesses an exact solution which takes
  the form of a logarithmic conformal field theory.  We show that the
  proper treatment of the locality conditions in the $SU(2)$ theory
  leads to the termination of the multifractal spectrum, or in other
  words to the termination of the infinite hierarchies of
  negative-dimensional operators that were thought to occur.  Based on
  arguments of logarithmic degeneracies, we conjecture that such a
  termination mechanism should be present for general $SU(N)$.
  Moreover, our results lead to the conclusion that the previous replica
  solution of this problem yields incorrect results.
\end{abstract}


\sloppy
\par

\section{Introduction}

There now exists a great deal of evidence that the wavefunctions in
disordered systems near a localization-delocalization transition
exhibit multifractal behaviour.  In other words, the moments of the
density of states scale not according to one given fractal dimension
like in a simple fractal, but rather according to an infinite set of
scaling exponents.  Since the most interesting features of
multifractality are beyond the reach of perturbation theory, one needs
to employ non-perturbative approaches. In this context the exactly
solvable problem of Dirac fermions in a random vector potential in two
dimensions has proved to be a very instructive training ground.  This
model was first considered in \cite{LudwigPRB50} in connection with
the integer quantum Hall transition.  Recently exact results for the full
multifractal spectrum for Dirac fermions interacting with a random
magnetic field (i.e. an Abelian vector potential) have been obtained
from a Gaussian field theory in an ultrametric space
\cite{ChamonPRL77}, and reproduced by mapping to a random energy model
\cite{Castillo9706084}.

The non-Abelian random vector potential case, originally introduced in
\cite{NersesyanPRL72} as an effective model for low-energy excitations
around the nodes of a d-wave superconductor, has very recently appeared
in physical problems of a very different nature, namely in non-Hermitian
quantum mechanics \cite{Mudry9712103}.  There, it appears as the
effective theory of a system driven by a strong {\it imaginary} vector
potential in the limit of weak disorder.  Non-Hermitian quantum
mechanics describes such problems as anomalous diffusion in a random
media \cite{Igichenko} and statistical mechanics of flux lines in
superconductors \cite{Hatano}. The multifractality is observed in
impurity-averaged correlators of local moments of wavefunctions, for
which expressions for the non-Abelian part need to be computed.
Although these correlators were investigated in
\cite{MudryNPB466,CauxNPB466}, we wish to report some new and surprising
aspects of the problem that were overlooked.  An earlier presentation of
some initial results can be found in \cite{CauxPRL}.

The non-Abelian Random Vector Potential (RVP) has already been treated
by both standard approaches for disordered systems, namely the replica
and SUSY ones.  The SUSY treatment can be found in \cite{MudryNPB466},
while the replica treatment was put forward in \cite{NersesyanPRL72} and
carried through in \cite{CauxNPB466}.  It was first recognized in this
publication that the associated conformal theory should be of a special
class, namely a {\it logarithmic} one.  However, both of these lines of
investigation suffered from various setbacks.  First of all, the SUSY
approach seemingly generated an infinite series of operators with
negative conformal dimensions for $SU(N)$ randomness.  %
On the other hand, the replica approach gave a different physical
content, coming from the fact that the conformal blocks of the theory
were slightly different.  There was consequently an urgent need to
provide a better comprehension of this system by performing some more
careful and extensive investigations.

As was pointed out in  \cite{BernardNPB441}, the disorder averaging in
the non-Abelian RVP problem can be performed {\it without} using the
SUSY or replica approaches in the limit of vanishing frequency and
infinite disorder strength.  This limit corresponds to the conformal
limit of the theory, which was studied in \cite{MudryNPB466}.  The RVP
model thus is a {\it critical} disordered system, which makes it
theoretically very interesting in its own right.  In addition, the
logarithmic nature of the CFT involved calls for better understanding.

Usually, CFTs are synonymous with power-law dependence of physical
correlators.  Much work has been done about the unitary minimal models,
for which there exists a unitary finite-dimensional representation of
the Virasoro algebra.  However, recently it has become clear that {\it
  logarithmic} dependence can appear for models outside of this class.
The first instance in which logarithms were introduced in conformal
correlators is to be found in \cite{RozanskyNPB376}.  It was first
recognized in \cite{GurarieNPB410} that this type of logarithmic
behaviour was associated to non-diagonalizability of the Virasoro
operator.  Subsequent applications of logarithmic operators were found
in critical disordered models \cite{CauxNPB466,MaassaraniNPB489},
$c_{p,1}$ and $c_{p,q}$ models
\cite{GaberdielNPB477,FlohrIJMPA11,Kausch9510149}, critical polymers and
percolation \cite{FlohrIJMPA11,Kausch9510149,SaleurNPB382},
gravitationally dressed CFTs \cite{BilalNPB449}, two-dimensional
magnetohydrodynamic turbulence \cite{Rahimi,FlohrNPB482}, D-brane recoil
in string theory \cite{KoganPLB375,KoganPLB387,EllisIJMPA12}, and were
recently shown to correspond to a novel bulk excitation in the Quantum
Hall state \cite{GurarieNPB498}.  Further new developments about
logarithmic CFTs can be found for example in
\cite{CauxNPB489,Koganh9705240,Rahimih9707060,KhorramiNPB497,%
Flohrh9707090,Rohsiepe}.

This paper aims to resolve some important remaining issues related to
fermions interacting with a non-Abelian random vector potential in the
conformal limit.  Our main result is to show that the correct treatment
of the {\it full} conformal field theory that is obtained from the SUSY
or direct treatment (which coincide) leads to the termination of the
parabolic multifractal spectrum before the scaling exponents $\tau^*(q)$
reach their maximal value.  They were previously thought to follow this
parabolic law to arbitrarily negative values.

The paper is organized as follows.  In section II we introduce the model
and present the conformal limit.  In section III we manipulate the
theory at the critical point and recall how to average over disorder
without SUSY or replicas.  In section IV we explicitly construct the
free-field formulation of the theory for $N=2$, with which multipoint
correlators are to be calculated explicitly in section V.  Section VI
discusses the fusion rules and the operator algebra of the logarithmic
conformal field theory, which are shown to have great impact on the
scaling behaviour to be expected.  The treatment for $N>2$ is presented
in section VII together with a discussion about the extension of the
$N=2$ results to this domain.  In section VIII we discuss the
consequences of our results on the multifractal spectrum.  Section IX
discusses the consequences on the replica solution, and we finish with
conclusions in section X.

\section{The Model}

We consider $N$ species of Dirac fermions living in a 2+1-dimensional
space and interacting through a disordered vector potential $A_{\mu}$
belonging to an $su(N)$ algebra ${\cal A}$, to which they are coupled
minimally.  The disorder allows for hopping between the different
species.  Since the vector potential is time-independent, different
Matsubara frequencies do not couple, and can be treated independently by
a Euclidean two-dimensional theory with explicit frequency dependence.
In fact, for a given realization of the disorder, the partition function
is given by the fermionic path integral
\begin{eqnarray}
Z[A_{\mu}] = \int\!\! {\cal D}[\bar{\Psi},\Psi]\; e^{{-S[\Psi, \omega,
    A_{\mu}]}}  
\end{eqnarray}
with the Dirac action
\begin{eqnarray}
S[\Psi, \omega, A_{\mu}] = \int\!\! d^2 x\;  \bar{\Psi}(x) [{\bf I}
~\otimes \not \! \partial -i \omega + i \not \!\! A] \Psi(x)
\end{eqnarray}
(since we are in a two-dimensional Euclidean space, we take the
Pauli matrices as Dirac $\gamma$ matrices, {\it i.e.} $ \not \!\!\!\!
A = A_{\mu} \otimes \sigma^{\mu}, \mu =1,2; \bar{\Psi} = 
\Psi^{\dagger} \sigma^1$).  

Disorder-dependent single-particle Green's functions can then be 
represented by the path integral
\begin{eqnarray}
G(x, y; \omega; A_{\mu}) = \frac{-i}{Z[A_{\mu}]} \int\!\! {\cal
  D}[\bar{\Psi},\Psi]\;
\Psi (x) \bar{\Psi} (y) e^{-S[\Psi, \omega, A_{\mu}]} \label{greensfunction}
\end{eqnarray}

Physical quantities are obtained by performing the disorder averaging
procedure on products of Green's functions.  We use the 
distribution functional
\begin{eqnarray}
\ln P[A_{\mu}] = -\frac{1}{\bar{g}} \int d^2 x ~\mbox{Tr} A_{\mu} (x)
A_{\mu} (x) 
\label{distribution} 
\end{eqnarray}
representing the usual $\delta$-correlated Gaussian white noise for 
the random vector potential.  

The impurity-averaged Green's function then reads
\begin{eqnarray}
G(x,y,\omega) = \overline{G(x, y; \omega; A_{\mu})} = \int {\cal D}
A_{\mu} G(x, y; \omega; A_{\mu}) P[A_{\mu}] = -i \int {\cal D} A_{\mu}
\frac{P[A_{\mu}]}{Z[A_{\mu}]} \int\!\! {\cal D}[\bar{\Psi},\Psi]\;
\Psi (x) \bar{\Psi} (y) e^{-S[\Psi, \omega, A_{\mu}]}
\end{eqnarray}

In the limits of infinite disorder strength $\bar{g} \to \infty$ and 
of vanishing frequency $\omega \to 0$, the theory becomes conformally 
invariant.  We will formulate and solve for correlation functions at this
conformal point, and later use renormalization group arguments to infer
the scaling behaviour of physical quantities like the density of states
and its local moments away from criticality.  We will use the notation
$G(x_1,...,x_n)$ to denote the impurity-averaged $n$-point correlator
at zero frequency.

\section{The Theory at the Critical  Point}

In the following sections, we will be concerned with the step-by-step
formulation of the theory at the conformally invariant critical point,
{\it i.e.}, for infinite disorder strength and vanishing frequency.  The
principles of this approach have been formulated by Bernard
\cite{BernardNPB441} and Mudry {\it et. al.}\cite{MudryNPB466}.  In this
work we show that the path integral is perfectly well-defined and
negative-dimensional operators do not appear in the physical operator
OPEs.

We start by separating the fermionic action into chiral parts:
\begin{eqnarray}
&&S[\Psi, A_{\mu}] = S_{+}[\Psi, A_+] + S_{-}[\Psi, A_-]
\nonumber \\
&&S_{\pm} [\Psi, A_{\pm}] = \int d^2 x \Psi_{\pm}^{\dagger} (x)
[\partial_{\pm} + i A_{\pm}(x) ] \Psi_{\pm} (x) 
\end{eqnarray}
where we have used the holomorphic and antiholomorphic derivatives and
fields $(2\partial = \partial_-, 2\bar{\partial} = \partial_+)$
\begin{eqnarray}
\partial_{\pm} = \partial_1 \pm i \partial_2 \nonumber \\
A_{\pm} = A_1 \pm i A_2 
\end{eqnarray}
where now $A_{\pm} \in su^C(N)=sl(N,{\Bbb C})$, the complex extension
${\cal A}^C$ of ${\cal A}$.  This translates, at the level of the path
integration over the vector fields, into transforming the double
integral of $A_i$ over $su(N)$ into a single integral over $su^C(N)$.
The non-compact nature of the $SU^C(N)$ group manifold, in contrast to
$SU(N)$, will be of crucial importance later on.

We parametrize the vector fields by fields $g_{\pm}$ belonging to the
complex extension $G^C$ of the group $G = SU(N)$: 
\begin{eqnarray}
A_{\pm} (x) = i \partial_{\pm} g_{\pm}(x) g_{\pm}^{-1}(x) \label{change}
\end{eqnarray}
The reality condition $A_+^{\dagger} (x) = A_- (x)$ translates into
$g_+^{\dagger} (x) = g_-^{-1} (x)$.  From now on, we will use the
notation $g_+ (x) = g (x)$.

Let us study, for a little while, the problem at fixed disorder.  We
can observe that the transformations 
\begin{eqnarray}
\Psi_{\pm} (x) \to g_{\pm} (x) \Psi_{\pm}^{\prime} (x),  ~~~~~~
\Psi_{\pm}^{\dagger} (x) \to {\Psi_{\pm}^{\prime}}^{\dagger} (x)
g_{\pm}^{-1} (x)  \label{fermiondecoupling}
\end{eqnarray}
completely decouple the fermions from the random vector potential.
It maps the minimally coupled Dirac action for $\Psi$ fermions into
the free Dirac action for $\Psi^{\prime}$ fermions, i.e.
\begin{eqnarray}
S[\Psi, A_{\mu}] \to S[\Psi^{\prime}] = \int d^2 x
[{\Psi^{\prime}}^{\dagger}_+ \partial_+ \Psi^{\prime}_+ +
{\Psi^{\prime}}^{\dagger}_- \partial_- \Psi^{\prime}_-  ]
\end{eqnarray}
generating the correlators
\begin{eqnarray}
\langle \Psi^{\prime}_+ (z_1) {\Psi^{\prime}}_+^{\dagger}(z_2) \rangle = 
\frac{1}{2\pi z_{12}}
\nonumber \\
\langle \Psi^{\prime}_- (\bar{z}_1) {\Psi^{\prime}}_-^{\dagger}
(\bar{z}_2) \rangle =
\frac{1}{2 \pi \bar{z}_{12}} \label{fermionope}
\end{eqnarray}
The Jacobian for this transformation,
\begin{eqnarray}
\frac{{\cal D}[\Psi,\bar{\Psi}]}{{\cal D}[\Psi^{\prime},\bar{\Psi}']} = 
\frac{\mbox{Det}[\not \! \partial + i \not \!\! A]}{\mbox{Det}[\not \!
  \partial]} 
\propto Z[A] 
\end{eqnarray}
has the very important property of being proportional to the partition
function at fixed disorder (up to the anomalous determinant
$\mbox{Det}[\not \!\! \partial]$, which is irrelevant for computation of
correlation functions but contributes to the total central charge), thus
cancelling it when computing the correlations for fixed disorder, {\it
  i.e.},
\begin{eqnarray}
G(x_1,...) = \int {\cal D} A_{\mu} \frac{1}{Z[A]} \int {\cal
D}[\Psi,\bar{\Psi}] \Psi... e^{-S[\Psi, A]} = \int {\cal
D} A_{\mu} \int {\cal
D}[\Psi^{\prime},\bar{\Psi}']\: g \Psi^{\prime}
... e^{-S[\Psi^{\prime}]}
\end{eqnarray}
This removes the need to invoke either the replica
or supersymmetry methods to perform explicitly the disorder averaging.
This also implies that we are not dealing with a free energy functional,
and that disconnected diagrams do appear in a perturbation expansion.

The original disorder averaging procedure involved two $su(N)$
algebra integrations (for the two components of the vector field).  We
transform these two integrations into an integration over a ``gauge''
sector and a ``coset'' sector, by writing $A_{\pm} = A^{(gauge)}_{\pm}
+ A^{(coset)}_{\pm}$.  The change of
variables (\ref{change}) transforms these two integrals in a single
group integration over the complex extension $SU^C(N)$, which we can
again separate into ``gauge'' and ``coset'' integrations by writing
$g_{\pm} = u H_{\pm}, u \in SU(N), H_{\pm} \in SU^C(N)/SU(N)$ (the
reality condition states then that $H_-^{-1} = H_+^{\dagger}$).  This
procedure induces a non trivial Jacobian in the path integral:
\begin{equation}
{\cal D} A_{\mu} = {\cal D} A_1 {\cal D} A_2 = {\cal D} A^{(gauge)} {\cal D}
A^{(coset)} = {\cal D} A^{(gauge)} {\cal D} H e^{2N W
[H^{\dagger} H]}  \label{jacobian} 
\end{equation}
where ${\cal D} H$ is the integration measure over the coset $SU^C(N)/SU(N)$,
and $N$ corresponds to the dual Coxeter number of $SU(N)$.  $W[h]$ is the
WZNW functional ($\partial B=S$) for the field $h \equiv H^{\dagger}H$
on level $k = -2N$ \cite{BernardNPB441,MudryNPB466} (note that the
topological term then has positive sign): 
\begin{eqnarray}
W [h] = \frac{1}{8 \pi} \int_{S} d^2 x \mbox{Tr}[\partial_{\mu} h
\partial_{\mu} h^{-1}] \quad + \frac{i}{12 \pi} \int_{B}\!\!  d^3x\; 
\epsilon_{\mu \nu \lambda} \mbox{Tr}[\partial_{\mu} h h^{-1} \partial_{\nu} h
h^{-1} \partial_{\lambda} h h^{-1} ]
\end{eqnarray}
Thus, the
partition function for $N$ species of Dirac fermions interacting with an
$su(N)$ random vector potential has completely factorized in the product
of two independent sectors: a free fermion (disorder independent) part,
and an $SU^C(N)$ part in which all disorder dependence is concentrated.

\section{Coset Factorization and Free Field Formulation of the WZNW Model}

The important point to notice from the previous formulas is that, since
the WZNW model depends only on the combination $H^{\dagger}H$ which is
invariant under left-multiplication by $u \in SU(N)$, i.e. $H \to u H $,
it is advantageous to proceed to the factorization of $SU^C(N)$ into the
subgroup $SU(N)$ and the coset $SU^C(N)/SU(N)$.  This can be explicitly
performed with the help of the Iwasawa decomposition of $SU^C(N)$
\cite{Helgason84}.  The Haar measure on $SU^C(N)$ then factorizes into the
product of the Haar measures on $SU^C(N)/SU(N)$ and $SU(N)$.

For simplicity, we specialize now to $N=2$.  Our derivation follows
essentially the approach to WZNW models as free field theories to be
found in \cite{GerasimovIJMPA5}.  For $g \in SU^C(2)$, the Iwasawa
decomposition takes the form
\begin{eqnarray}
g = u \left( \begin{array}{cc} 
e^{\phi /2} & 0 \\
0 & e^{-\phi /2} 
\end{array} \right) \left( \begin{array}{cc}
1 & \mu_+ \\ 0 & 1 
\end{array} \right)
\label{Iwasawa}
\end{eqnarray}
with $u \in SU(2)$ and $\phi \in {\Bbb R}$, $\mu_+ \in {\Bbb C}$.  The
integral measure for this parametrization is then given by (see
derivation in the Appendix)
\begin{eqnarray} 
d g = d u\; e^{2\phi} d \phi d \mu_+ d \mu_-
\label{haarmeasure} 
\end{eqnarray}
$du$ being the Haar measure on $SU(2)$.

The field appearing in the Jacobian (\ref{jacobian}), being $SU(2)$
independent, can be expressed in terms of the fields $(\phi, \mu_+,
\mu_-)$ as 
\begin{eqnarray}
h = H^{\dagger} H = \left( \begin{array}{cc}
e^{\phi} &~ \mu_+ e^{\phi} \\
\mu_- e^{\phi} &~ e^{-\phi} + \mu_+ \mu_- e^{\phi} 
\end{array} \right)
\end{eqnarray}
Using this parametrization in the Jacobian (\ref{jacobian}) gives
\begin{eqnarray}
kW[h] = \frac{-k}{4 \pi} \int d^2 x [ (\partial_{\mu}
\phi)^2 + e^{2\phi} \partial_-\mu_+ \partial_+ \mu_-]
\end{eqnarray}
as the action for the $SU^C(2)/SU(2)$ fields $(\phi, \mu_+,
\mu_-)$.  For negative $k$, this action is positive definite.  Once
again:  even though the Jacobian (\ref{jacobian}) induces a WZNW
model with a negative level, which is not well-defined at the level
of the path integral for a group manifold with a positive metric,
the fact that the WZNW field $h$ belongs to the coset $SU^C(2)/SU(2)$
which has a negative-definite metric makes the coset action a
positive-definite functional.  This is reminiscent of the coset
constructions on non-compact manifolds to be found in \cite{BarsNPB334},
where the coset formed of the non-compact group and its maximally
compact subgroup forms a well-defined unitary conformal theory when
one considers a negative level for the underlying Kac-Moody algebra.   

The WZNW model has the property of being invariant with respect to
Kac-Moody chiral current algebras.  The Kac-Moody currents, given by 
(for negative level)
\begin{eqnarray}
J(z) = -k \partial h h^{-1}, ~~~~~ \bar{J}(\bar{z}) = -k
h^{-1} \bar{\partial} h 
\end{eqnarray}
have, in terms of the new fields, the representation
\begin{eqnarray}
&&J(z) = J^+ \sigma^- + J^- \sigma^+ + H \sigma^3 \nonumber \\ \\
&&J^+ = -k [ 2 \mu_- \partial \phi + \partial \mu_- -
\mu_-^2e^{2\phi}\partial \mu_+] \nonumber \\
&&J^- = -k e^{2\phi} \partial \mu_+ \nonumber \\ 
&&H = -k [\partial \phi - \mu_- e^{2\phi}\partial \mu_+ ] \nonumber \\ \\
&&\bar{H} = H^{\dagger} ~~~ \bar{J}^{\pm} = {J^{\mp}}^{\dagger}
\label{currents1} 
\end{eqnarray}

In the path integration over $(\phi, \mu_+, \mu_-)$, the presence of
the factor $e^{2\phi}$ in the measure as well as the presence of the
combination $e^{2\phi} 
\partial \mu_+$ in the expression for the current $J(z)$ make the
change of variables
\begin{eqnarray}
\omega_- (z) = -ke^{2\phi} \partial \mu_+ \label{changeofvar}
\end{eqnarray}
very convenient, since it brings us to a path integration over
free fields.

This and the properly regularized anomalous determinant associated to
this transformation \cite{GerasimovIJMPA5}
\begin{eqnarray}
\mbox{Det}[e^{2\phi}\partial] \to (\mbox{Det}[e^{-2\phi} \bar{\partial}
e^{2\phi} \partial])^{1/2} 
\end{eqnarray}
which shifts the action for $\phi$ by
\begin{eqnarray}
\frac{1}{4\pi} \int d^2x [2 (\partial_{\mu} \phi)^2 + {\cal
R} \phi] 
\end{eqnarray}
modify the coset action to
\begin{eqnarray}
S[\phi, \omega_-, \mu_-] = \frac{1}{4\pi} \int d^2 x
[(-k-2)(\partial_{\mu} \phi)^2 + {\cal R} \phi] + \frac{1}{2\pi} \int
d^2 x \omega_- \partial_+ \mu_- 
\end{eqnarray}
i.e. the anomaly shifts $k \to k +2 \equiv -q^2$ (recall that in the
given case $k = - 4, q^2 = 2$) in front of $\phi$
and introduces the 
Riemann curvature ${\cal R}$ of the manifold in the action for $\phi$,
which will change its conformal properties.   
By rescaling $\phi \to \frac{1}{\sqrt{2}q} \phi^{\prime}$, the action
and Operator Product Expansions for our free fields read 
\begin{eqnarray}
S[\phi^{\prime}, \omega_-, \mu_-] = \frac{1}{8\pi} \int d^2x
[(\partial_{\mu} \phi^{\prime})^2 + \frac{\sqrt{2}}{q}{\cal R} 
\phi^{\prime}] + \frac{1}{2\pi} \int d^2 x \omega_- \partial_+ \mu_-
\nonumber  \\ \\
\phi^{\prime}(z_1) \phi^{\prime}(z_2) = -\ln z_{12}, ~~~
\bar{\phi}^{\prime}(\bar{z}_1) \bar{\phi}^{\prime}(\bar{z}_2) = -\ln
\bar{z}_{12}, ~~~\mu_-(z_1) \omega_-(z_2) = \frac{1}{z_{12}}
\end{eqnarray}
in which we have performed the separation of $\phi^{\prime}$ in terms
of holomorphic and anti-holomorphic parts $\phi^{\prime}(z, \bar{z}) =
\phi(z) + \bar{\phi}(\bar{z})$, valid at the level of correlation
functions.

The proper expression for the currents in terms of our free fields is
then given by (\ref{currents1}), but only after performing the same
transformations as above on $\phi$, i.e. the shift $k\phi \to
(k+2)\phi$ and the rescaling $\phi \to \frac{1}{\sqrt{2}q} \phi^{\prime}$:
\begin{eqnarray}
&&J^- (z) = \omega_- (z) \nonumber \\
&&J^+ (z) = \sqrt{2} q \mu_- (z) \partial
\phi^{\prime} (z) - k \partial \mu_- (z) - \mu_-^2 (z) \omega_-(z)
\nonumber \\
&&H (z) = \frac{1}{\sqrt{2}} q \partial \phi^{\prime}(z) - \mu_- (z)
\omega_- (z) \label{currents2}
\end{eqnarray}
In all of these, the usual normal-ordering procedure is implied.
The currents have the correct OPEs for generators in the Cartan-Weyl
basis of $SU(2)_k$, i.e.
\begin{eqnarray}
&&H (z_1) J^{\pm} (z_2) = \frac{\pm 1}{z_{12}} J^{\pm} (z_2) + ...\nonumber
\\
&&J^+ (z_1) J^- (z_2) = \frac{k}{z_{12}^2} + \frac{2}{z_{12}} H (z_2)
+ ... \nonumber \\
&&H(z_1) H(z_2) = \frac{k/2}{z_{12}^2} + ... \label{currentopes}
\end{eqnarray}
Thus, what we have done is the following:  by using the Iwasawa
decomposition (\ref{Iwasawa}), we have explicitly shown that the
level $k = -2N = -4$ coset $SU^C(2)/SU(2)$ possesses a Wakimoto free
field representation with Cartan-Weyl generators carrying an affine
$SU(2)$ current algebra with level analytically continued to negative
value. 

The stress-energy tensor
\begin{eqnarray}
T(z) = \frac{-1}{q^2} [\frac{1}{2}(J^+ J^- + J^- J^+) + H^2]
\end{eqnarray}
is, in terms of our free fields,
\begin{eqnarray}
T(z) = -\frac{1}{2}\partial \phi^{\prime}\partial
\phi^{\prime} - \frac{1}{\sqrt{2}q} \partial^2 \phi^{\prime} -
\omega_- \partial \mu_- 
\end{eqnarray}
The central charge is readily checked to be $c= 3k/(k+2)$ as for
the normal case of positive $k$ (in our case, $c=6$).  
The $\phi^{\prime}$ part of $T(z)$ is that of
a Dotsenko-Fateev free field with imaginary background charge $\alpha_0 =
i/(2q)$.  Vertex operators $V_{\alpha} (z) = e^{\alpha
\phi^{\prime}(z)}$ then have holomorphic conformal weight $h_{\alpha} =
-(\alpha/2) (\alpha +\sqrt{2}/{q})$.

Primary operators $V_{\beta}$ can be found by requiring the correct
OPEs with $T(z)$ and the generators of the Kac-Moody algebra
(\ref{currents2}).  A highest-weight finite-dimensional representation
is available, composed of the $2j+1$ operators $V_{jm}(z)$ given by
\begin{eqnarray}
V_{jm}(z) = e^{\frac{\sqrt{2}j}{q}\phi^{\prime}(z)} [-\mu_-(z)]^{j+m}
\end{eqnarray}
for which the OPEs with the currents have the correct $SU(2)$ form
\begin{eqnarray}
&&J^{\pm}(z_1) V_{jm}(z_2) = \frac{j \mp m}{z_{12}} V_{j m \pm 1}(z_2)
\nonumber \\
&&H(z_1) V_{jm}(z_2) = \frac{m}{z_{12}} V_{jm}(z_2) 
\end{eqnarray}
These primary operators all possess identical conformal weights
\begin{eqnarray}
h_{jm} = -\frac{\alpha}{2}(\alpha + \frac{\sqrt{2}}{q}) = -
\frac{j(j+1)}{q^2} = \frac{j(j+1)}{k+2} 
\end{eqnarray}
(note that $\mu_-$ is a weight zero operator, as can be seen from its
OPE with $T(z)$, and consequently does not contribute to the weight of
$V_{jm}$).  Weights are thus negative in any spin-$j$ representation
for our value $k=-4$.

To calculate correlation functions, we follow the general procedure
outlined in \cite{DotsenkoNPB}.  This starts with the
introduction of dual 
operators $\tilde{V}_{jm}$, generalizing the usual dual vertex
operators of the Coulomb-gas formalism.  The highest-weight
operator $\tilde{V}_{j-j}$ is given by 
\begin{eqnarray}
\tilde{V}_{j-j} (z) = e^{-\frac{\sqrt{2}}{q}(3+j)\phi^{\prime}}(z)
\omega_-^{3+2j} (z) \label{simpleform}
\end{eqnarray}
Acting on this with $J^+$ produces the other states in the
representation, but there is a slight setback:  they do not have such
a simple form as (\ref{simpleform}).  
To obtain a correlator involving another dual state than
(\ref{simpleform}), the trick is to use the conformal Ward identities
and derive it from another correlator involving  (\ref{simpleform}).
For future reference, we here give the two dual
operators in the fundamental spin-$1/2$ representation:
\begin{eqnarray}
&&\tilde{V}_-(z) \equiv \tilde{V}_{1/2, -1/2} =
e^{-\frac{7}{\sqrt{2}q}\phi^{\prime}(z)} \omega_-^4 (z) \nonumber \\
&&\tilde{V}_+(z) \equiv \tilde{V}_{1/2, 1/2} = \left[4\sqrt{2} q \partial
\phi^{\prime} (z) \omega_-^3(z) - \mu_-(z) \omega_-^4(z) -12\partial
\omega_-(z) \omega_-^2 (z)\right] e^{-\frac{7}{\sqrt{2}q}\phi^{\prime}(z)}
\end{eqnarray}

The general prescription for the calculation of correlators
\cite{DotsenkoNPB} then requires that we replace one of the
fields by its dual, and use screening charges if necessary to satisfy
the neutrality conditions, which are now modified to 
\begin{eqnarray}
\sum \alpha_i = -\frac{6}{\sqrt{2}q} = -3, ~~~~~ N_{\omega_-} -
N_{\mu_-} = 3 \label{neutrality}
\end{eqnarray}
with $\alpha$ understood as the numerical coefficient of the vertex
operator, used as in $e^{\frac{\alpha}{\sqrt{2}q}\phi^{\prime}}$, and
where $N_{\omega_-}$ and  $N_{\mu_-}$ count these respective operators
in the correlator.  Note that this last condition is equivalent to the
requirement that the correlator be a singlet state, e.g. that it
contains an equal number of $+$ and $-$ indices in the spin-$1/2$
case.  

The last ingredient we will need is the screening operator.  It can be
determined in two ways:  first, it must commute with all current
generators up to a total differential; but also, it can be found from
the change of variables (\ref{changeofvar}).  Inverting this, we can
see that over every noncontractable cycle on the Riemann surface, we
must have  
\begin{eqnarray}
\oint d \zeta \omega_-(\zeta) e^{-\frac{\sqrt{2}}{q}\phi^{\prime}
(\zeta)} = -k \oint d \mu_+ = 0
\end{eqnarray}
Enforcing this condition can be done by putting in the path integral
the (infinite) product, over all closed cycles, of $\delta$ functions
\begin{eqnarray}
\prod_c \delta (\oint \omega_- e^{-\frac{\sqrt{2}}{q}\phi^{\prime}}) =
\int d \lambda \exp (i\lambda \oint \omega_-
e^{-\frac{\sqrt{2}}{q}\phi^{\prime}}) 
\end{eqnarray}
Expanding in powers of $\lambda$ then produces the screening charge
insertions, which we will denote by the operator ${\cal Q} (z)$.

We have explicitly constructed the holomorphic operators of the
theory.  An identical procedure can be done for the antiholomorphic
part, using instead of (\ref{changeofvar}) the conjugate transformation
\begin{eqnarray}
\omega_+ (\bar{z}) = -k e^{2\phi}\bar{\partial}\mu_-
\end{eqnarray}
This way, primary highest-weight states of the antiholomorphic
currents can be constructed, as well as
their duals, and a conjugate screening charge can be found.  We will denote
all of these by $\bar{V}$, etc.  

The matrix field $h(z, \bar{z})$ can then be expressed, in correlation
functions, in terms of the tensor product of left and right primary
spin-$1/2$ fields as 
\begin{eqnarray}
h(z, \bar{z}) = \left( \begin{array}{cc} 
V_-(z)\bar{V}_-(\bar{z}) & V_-(z) \bar{V}_+(\bar{z}) \\
V_+(z) \bar{V}_-(\bar{z}) & V_+(z)\bar{V}_+(\bar{z}) \end{array}
\right) \label{hvertex}
\end{eqnarray}

The two-point functions of primary fields 
\begin{eqnarray}
\langle V_{\pm}(z_1) \tilde{V}_{\mp} (z_2) \rangle = z_{12}^{3/4}
\end{eqnarray}
give for the coset two-point correlator
\begin{eqnarray}
\langle h_{a_1b_1}(z_1, \bar{z}_1) h^{-1}_{b_2 a_2}(z_2, \bar{z}_2)
\rangle = \delta_{a_1 a_2} \delta_{b_1 b_2} |z_{12}|^{3/2}
\label{twopoint}
\end{eqnarray}

We have thus seen how coset correlations can be simply obtained by
calculating the relevant conformal blocks of our primary fields.  This
will be done for the four-point function in the following section.

\section{Multi Point  Correlation Functions}

Let us start by moving back to our original physical problem, and see
what we are now capable of doing.  By considering the limiting case
$\bar{g} \to \infty$, we have decomposed our original fermions into
free fermions $\psi^{\prime}$ together with $SU^C(N)/SU(N)$ coset
operators $h$.  Thus, by considering correlators of the local operators
\begin{eqnarray}
&{\cal O}(z, \bar{z}) = {\Psi_-^{\dagger}}_a {\Psi_+}_a &=
{\Psi_-^{\prime}}^{\dagger}_a h_{ab} {\Psi^{\prime}_+}_b \nonumber \\
&{\cal O}^{-1} (z, \bar{z}) = {\Psi_+^{\dagger}}_a {\Psi_-}_a &=
{\Psi_+^{\prime}}^{\dagger}_a h^{-1}_{ab} {\Psi^{\prime}_-}_b
\label{localoperators}
\end{eqnarray}
we explicitly sum over the $SU(N)$ indices, leaving only the coset
operators once the decoupling transformation (\ref{fermiondecoupling})
has been done.  The $SU(N)$ path integration then simply factorizes
out of the effective generating functional. 

The two-point function
\begin{eqnarray}
\langle {\cal O}(1) {\cal O}^{-1}(2) \rangle \sim \frac{1}{|z_{12}|^2}
\langle \mbox{Tr}[h(1) h^{-1}(2)]\rangle  \sim \frac{1}{|z_{12}|^{1/2}}
\label{twopointO} 
\end{eqnarray}
shows that the operator ${\cal O}$ has conformal weights $(1/8, 1/8)$
(i.e. a spinless operator of dimension $1/4$).  This reproduces the
previously known result \cite{NersesyanPRL72}.

Let us now turn to the four-point function.  By performing the
fermionic contractions, we get 
\begin{eqnarray}
\langle {\cal O}(1) {\cal O}^{-1}(2) {\cal O}(3) {\cal O}^{-1}(4)
\rangle \quad \sim \frac{1}{|z_{12}z_{34}|^2} \langle
\mbox{Tr}[h(1)h^{-1}(2)] \mbox{Tr}[h(3) h^{-1}(4)] \rangle + \nonumber
\\ \qquad + \frac{1}{z_{14}z_{23} \bar{z}_{12} \bar{z}_{34}} \langle
\mbox{Tr}[h(1) h^{-1}(2) h(3) h^{-1}(4)] \rangle \qquad + (2
\leftrightarrow 4)
\label{fourpoint} 
\end{eqnarray}
By using (\ref{hvertex}), we can see that all correlators appearing in
(\ref{fourpoint}) can be expressed in terms of the three vertex
correlators 
\begin{eqnarray}
C_1 = \langle V_+(z_1) V_-(z_2) V_+(z_3) V_-(z_4) \rangle \nonumber \\
C_2 = \langle V_+(z_1) V_+(z_2) V_-(z_3) V_-(z_4) \rangle \nonumber \\
C_3 = \langle V_-(z_1) V_+(z_2) V_+(z_3) V_-(z_4) \rangle \nonumber
\end{eqnarray} 
These can be calculated using the prescription of the last section:
we replace the operator in position $4$ by its dual (note that, in the
definitions above, we have performed an overall spin flip, which does
not change the correlators but allows to use the more simple
expression $\tilde{V}_-$), and insert the necessary screening charges
(here, only one).  We use the projection invariance to perform the
analytic mapping $(z_1, z_2, z_3, z_4) \to (0, z, 1, \infty), z =
\frac{z_{12}z_{34}}{z_{13}z_{24}}$, for which we get
\begin{eqnarray}
C_i (z_1...z_4) = [z_{13}z_{24}]^{3/4} C_i (z), ~~~~~C_i (z) =
\lim_{z_{\infty} \to \infty} z_{\infty}^{3/4} C_i (0, z, 1,
z_{\infty})
\end{eqnarray}
Performing the integrals (with $C_i (z) = [z(1-z)]^{3/4}
\tilde{C}_i (z)$) yields
\begin{eqnarray}
&&\tilde{C}_1^a (z) = F(3/2, 3/2; 2; z) \nonumber \\
&&\tilde{C}_1^b(z) = F(3/2, 3/2; 2; 1-z) \nonumber \\
&&\tilde{C}_2^a (z) = F(1/2, 3/2; 1; z) \nonumber \\
&&\tilde{C}_2^b(z) = \frac{1}{2} F(1/2, 3/2; 2; 1-z) \nonumber \\
&&\tilde{C}_3^a (z) = \frac{1}{2} F(1/2, 3/2; 2; z) \nonumber \\
&&\tilde{C}_3^b(z) = F(1/2, 3/2; 1; 1-z) \label{firstblocks}
\end{eqnarray}
in which the two independent solutions for each correlator are labeled
by the integration contours $a: (0,z), b: (z,1)$ used in the
screening.  The $b$ contour solutions have logarithmic behaviour near
$z=0$, whereas the $a$ ones have logarithmic behaviour near $z=1$.  We
are thus in the presence of a logarithmic conformal field theory,
whose first examples appeared in \cite{RozanskyNPB376,GurarieNPB410}.

The ``full'' correlator, in which the $\bar{z}$ dependence has been
included, has to be built by combining the conformal blocks is such a
way as to satisfy the crossing symmetry constraints, as well as the
requirement of single-valuedness.  Crossing symmetry is easily
satisfied in view of the extremely simple monodromy properties of the
solutions (\ref{firstblocks}).  The constraint of single valuedness
(for example, around $z=0$) forces the cross-multiplication of $a$ and
$b$ solutions respectively in the $z$ and $\bar{z}$ sectors, without
which terms of the form $\ln z \ln \bar{z}$ would appear, which
obviously are to be excluded in a single-valued correlator.  

In fact, by explicitly multiplying out (\ref{hvertex}) and using the
above solutions, we find, using the standard identities 
\begin{eqnarray} 
&& F(3/2, 3/2; 2; z) + \frac{1}{2} F(1/2, 3/2; 2; z) = \frac{3}{2}
(1-z) F(3/2, 5/2; 2; z) \nonumber \\ 
&& F(3/2, 3/2; 2; z) - F(1/2, 3/2; 1; z) = \frac{3}{8} z F(3/2, 5/2;
3; z) \nonumber \\ 
\end{eqnarray}
that the four-point correlator can be written as a function of the
conformal blocks
\begin{eqnarray}
&&\tilde{F}_1^a (z) = 2\sqrt{2} (1-z) F(3/2, 5/2; 2; z) \nonumber \\
&&\tilde{F}_1^b (z) = (1-z) F(3/2, 5/2; 3; 1-z) \nonumber \\
&&\tilde{F}_2^a (z) = z F(3/2, 5/2; 3; z) \nonumber \\
&&\tilde{F}_2^b (z) = 2\sqrt{2} z F(3/2, 5/2; 2; 1-z) \label{blocks}
\end{eqnarray}
These can be verified to be the conformal blocks that one would obtain
by solving straightforwardly the $SU(2)_{-4}$ Knizhnik-Zamolodchikov
equations.  We have obtained them using a somewhat different route,
through the coset parametrization.  Thus, even though naively the
$SU(2)_{-4}$ WZNW model seems ill-defined at the level of the path
integral (since the kinetic term, for a negative level, has the wrong
sign !), the fact that we are in reality working on the coset
$SU^C(2)/SU(2)$ which has been shown above to carry the same current
algebra makes the use of the KZ equations valid.  Even though we have
proved this in detail only for $N=2$, we can expect that the same will
be true for $N > 2$ as well.  One very important remark to make at this
point is that these conformal blocks do not reproduce the conformal
blocks obtained within the replica approach \cite{CauxNPB466}.  Although
the operator dimensions are correctly reproduced, the correlators and
OPEs are incorrect.  We will return to this point later on.

The full coset correlator 
\begin{eqnarray}
&& H_{a_1...b_4} (z_1, z_2, z_3, z_4) = |z_{13}z_{24}|^{3/2}
H_{a_1...b_4} (z, \bar{z}) \nonumber \\ 
&&=\langle h_{a_1 b_1} (z_1,\bar{z}_1) h_{b_2 a_2}^{-1} (z_2, \bar{z}_2)  
h_{a_3 b_3} (z_3, \bar{z}_3) h_{b_4 a_4}^{-1} (z_4, \bar{z}_4) \rangle 
\end{eqnarray}
can now be built as mentioned above, by solving the monodromy problem,
and insuring single-valuedness in the complex plane.  The correlator
can be projected as usual onto the singlets
\begin{eqnarray}
I_1 = \delta_{a_1 a_2} \delta_{a_3 a_4} ~~~~~~ I_2 = \delta_{a_1 a_4}
\delta_{a_2 a_3} \nonumber \\
H_{a_1...b_4} (z, \bar{z}) = \sum_{i,j = 1,2} I_i \bar{I}_j H_{ij} (z,
\bar{z}) 
\end{eqnarray}
with, by single-valuedness,
\begin{eqnarray}
H_{ij} (z, \bar{z}) = \alpha |z(1-z)|^{3/2} [\tilde{F}_i^a (z)
\tilde{F}_j^b (\bar{z}) + (a \leftrightarrow b)]
\end{eqnarray}
where $\alpha$ is some constant chosen to satisfy consistency with the
normalization of the two-point function.  

The full four-point correlator for the local operators
(\ref{localoperators}), with free fermion contribution, finally reads
\begin{eqnarray}
\langle {\cal O}(1)...{\cal O}^{-1}(4) \rangle && \sim
\left|\frac{z_{12}z_{34}z_{14}z_{23}}{z_{13}z_{24}} \right|^{3/2}
\left\{ \frac{P(I_1)}{z_{12}z_{34}} + \frac{P(I_2)}{z_{14}z_{23}}
\right\}
\left\{\frac{\bar{P}(\bar{I}_1)}{\bar{z}_{12}\bar{z}_{34}} +
\frac{\bar{P}(\bar{I}_2)}{\bar{z}_{14}\bar{z}_{23}} \right\}
H_{a_1...b_4} (z, \bar{z}) \label{fourpointfunction}
\end{eqnarray}
where $P$ and $\bar{P}$ are projectors onto the singlets,
i.e. $P(I_i)\bar{P}(\bar{I}_j) H_{a_1...b_4} = H_{ij}$.

 For further discussion we shall need the four-point correlator of the
physical field ${\cal M}$ (related to the local density of states),
which is composed of the sum of the two operators
(\ref{localoperators}):
\begin{eqnarray}
{\cal M}(z, \bar{z}) = \mbox{Tr}\bar{\Psi} \Psi = {\cal O}(z, \bar{z}) +
{\cal O}^{-1} (z, \bar{z}) \label{density}
\end{eqnarray} 
The correlator we are interested in is made up of the
different permutations of (\ref{fourpointfunction}) that appear when
we make use of (\ref{density}).  In fact, it reads
\begin{eqnarray}
&& \langle {\cal M} (1) {\cal M}(2) {\cal M}(3) {\cal M} (4) \rangle =
2 \langle {\cal O} (1) {\cal O}^{-1} (2)  {\cal O} (3) {\cal O}^{-1}
(4) \rangle + (2\leftrightarrow 3) + (3 \leftrightarrow 4) 
\end{eqnarray}
The permutation $2 \leftrightarrow 3$ maps $z$ to $1/z$, whereas $3
\leftrightarrow 4$ maps it to $z/(z-1)$.  
The analytic continuation of
the conformal blocks (\ref{blocks}) to these regions can be performed
directly by using hypergeometric function identities, or more easily
by modifying the original screening charge integration contours
appropriately.  We find
\begin{eqnarray}
&\tilde{F}_1^a (1/z) =& - z^{3/2} \tilde{F}_1^a (z) \nonumber \\
&\tilde{F}_1^b (1/z) =& - z^{3/2} \tilde{F}_1^b (z) \nonumber \\
&\tilde{F}_2^a (1/z) =& - 4 z^{3/2} F(1/2, 5/2; 2; z) \nonumber \\
&\tilde{F}_2^b (1/z) =& 2\sqrt{2}  z^{3/2} F(1/2, 5/2; 2; 1-z) 
\label{analyticcontinuations1/z}
\end{eqnarray}
and 
\begin{eqnarray}  
&\tilde{F}_1^a (\frac{z}{z-1}) =&  2\sqrt{2} (1-z)^{3/2} F(1/2, 5/2;
2; z) \nonumber \\
&\tilde{F}_1^b (\frac{z}{z-1}) =&  -4 (1-z)^{3/2} F(1/2, 5/2; 2; 1-z)
\nonumber \\
&\tilde{F}_2^a (\frac{z}{z-1}) =&  -(1-z)^{3/2} \tilde{F}_2^a (z)
\nonumber \\
&\tilde{F}_2^b (\frac{z}{z-1}) =&  -(1-z)^{3/2} \tilde{F}_2^b (z)
\label{analyticcontinuationsz/z-1} 
\end{eqnarray}
These, together with (\ref{fourpointfunction}) and
(\ref{analyticcontinuations1/z}), allow us to write the exact
expression for the full four-point ${\cal M}$ correlator in our model:

\begin{eqnarray}
\langle {\cal M}(1) {\cal M}(2) {\cal M}(3) {\cal M}(4) \rangle \sim
\frac{1}{|z_{12}z_{34}|^{1/2}} |1-z|^{3/2}  \left[ {\cal H}_{11} +
\frac{\bar{z}}{1-\bar{z}} {\cal H}_{12} + \frac{z}{1-z} {\cal H}_{21}
+ \frac{|z|^2}{|1-z|^2} {\cal H}_{22} \right] \label{raw}
\end{eqnarray}
where
\begin{eqnarray}
{\cal H}_{11} (z, \bar{z}) &&= H^{\prime}_{11}(z, \bar{z}) + |z|^{-1}
H^{\prime}_{11} (\frac{1}{z}, \frac{1}{\bar{z}}) + |1-z|^{-3}
H^{\prime}_{11} (\frac{z}{z-1}, \frac{\bar{z}}{\bar{z}-1}) \nonumber
\\ 
{\cal H}_{12} (z, \bar{z}) &&= H^{\prime}_{12}(z, \bar{z}) - z
|z|^{-3} H^{\prime}_{12} (\frac{1}{z}, \frac{1}{\bar{z}}) -
(1-\bar{z}) |1-z|^{-3} H^{\prime}_{12} (\frac{z}{z-1},
\frac{\bar{z}}{\bar{z}-1}) \nonumber \\ 
{\cal H}_{21} (z, \bar{z}) &&= H^{\prime}_{21}(z, \bar{z}) - \bar{z}
|z|^{-3} H^{\prime}_{21} 
(\frac{1}{z}, \frac{1}{\bar{z}}) - (1-z)
|1-z|^{-3} H^{\prime}_{21} (\frac{z}{z-1}, \frac{\bar{z}}{\bar{z}-1})
\nonumber \\ 
{\cal H}_{22} (z, \bar{z}) &&= H^{\prime}_{22}(z,
\bar{z}) + |z|^{-3} H^{\prime}_{22} (\frac{1}{z}, \frac{1}{\bar{z}})
+ |1-z|^{-1} H^{\prime}_{22} (\frac{z}{z-1},
\frac{\bar{z}}{\bar{z}-1})
\end{eqnarray}
with the correlators $H^{\prime}_{ij}$ given by
\begin{eqnarray} 
&&H^{\prime}_{ij}(z, \bar{z}) =  \left[ \tilde{F}_i^a (z)
\tilde{F}_j^b(\bar{z}) + \tilde{F}_i^b (z) \tilde{F}_j^a
(\bar{z}) \right] \label{completefourpointfunction}
\end{eqnarray}
  From Eq.(\ref{raw}) we extract the OPE of ${\cal M}$ with
itself, which in turn determines the scaling of products of local
densities of states.

\section{Fusion Rules and Operator Algebra}

We will start here by studying the fusion rules of coset operators
$h(z, \bar{z})$.  There are two cases of interest:  the fusion $h
h^{-1}$, and $h h$.  These two cases can be obtained from the coset
correlator 
\begin{eqnarray}
&&\langle h_{a_1 b_1} (z_1,
\bar{z}_1) h_{b_2 a_2}^{-1} (z_2, \bar{z}_2)  
h_{a_3 b_3} (z_3, \bar{z}_3) h_{b_4 a_4}^{-1} (z_4, \bar{z}_4) \rangle
\sim \left|\frac{z_{12}z_{34}
    z_{14}z_{23}}{z_{13}z_{24}}\right|^{3/2} 
\sum_{i,j} I_i \bar{I}_j \left[ \tilde{F}_i^a (z) \tilde{F}_j^b
  (\bar{z}) + a \leftrightarrow b \right]
\end{eqnarray}
by respectively considering the limits $z_1 \to z_2$ in the above
correlator, and in the one with $2$ and $3$ permuted.  These
correspond to taking $z \to 0$ in the appropriate solutions, which
involves in the second case the analytic continuations $z \to 1/z$.  

Using the formulas contained in the appendix, we find
that the dominant contributions come from $H_{11}$, associated with
the $I_1, \bar{I}_1$ singlets (in the second case, this means
contracting $1$ with $3$, $2$ with $4$).  This yields (we here omit for
simplicity the index structure)
\begin{eqnarray}
h (z_1, \bar{z}_1) h^{-1} (z_2, \bar{z}_2) \sim 
|z_{12}|^{3/2} \left[ 
\frac{4/3}{z_{12}} A(z_2) +\frac{4/3}{\bar{z}_{12}} \bar{A}(\bar{z}_2)
+  4 {\cal I} + 2 \tilde{D}  
(z_2, \bar{z}_2) + \ln |z_{12}| \tilde{C} (z_2, \bar{z}_2)  \right]
\label{opehh-1}
\end{eqnarray}
in which ${\cal I}$ is the identity operator.  The correlators of the
operators appearing in (\ref{opehh-1}) are
\begin{eqnarray}
&&\langle A(z_1) A(z_2) \rangle \sim z_{12}^2;\quad \langle
\bar{A}(\bar{z}_1) \bar{A} (\bar{z}_2) \rangle \sim \bar{z}_{12}^2
\nonumber \\ \\
&&\langle \tilde{D} (z_1, \bar{z}_1) \tilde{D} (z_2, \bar{z}_2)
\rangle \sim - c_1 - \ln |z_{12}| \nonumber \\
&&\langle \tilde{D} (z_1, \bar{z}_1) \tilde{C} (z_2, \bar{z}_2)
\rangle \sim 1 \nonumber \\
&&\langle \tilde{C} (z_1, \bar{z}_1) \tilde{C} (z_2, \bar{z}_2) \rangle = 0
\end{eqnarray}
where $c_1$ is some constant, unimportant for our purposes.

Some comments are in order here.  We notice that the most relevant
operators appearing in the OPE (\ref{opehh-1}), $A(z)$ and
$\bar{A}(\bar{z})$, possess conformal weights $(-1, 0)$ and $(0, -1)$
respectively.  Usually, the fusion rules for WZNW models would imply
that the adjoint operator, whose conformal weights are 
\begin{eqnarray}
h_{ad} = \bar{h}_{ad} = \frac{N}{N+k} = -1
\end{eqnarray}
should appear in the OPE (\ref{opehh-1}).  But the term in the
four-point function pointing to the presence of such an operator,
corresponding to a term $\sim 1/|z|^2$ inside the brackets of
(\ref{opehh-1}), does not appear since we
are not allowed, by the requirement of single valuedness, to multiply
the $b$ contour solutions in the holomorphic and antiholomorphic
sectors together.  This requirement cannot be seen from the chiral
conformal algebra: it comes out of the solution for the full correlator.

Thus, single valuedness in the logarithmic conformal field theory that
we are considering here changes the scaling behaviour expected from
considerations not taking it into account, as in \cite{MudryNPB466}.  In
fact, the operators (\ref{localoperators}) exactly correspond to the
local fields considered there.

We can now very easily derive the OPE of ${\cal O}
{\cal O}^{-1}$.  It has the exact same logarithmic form as above, with
the exception that the proper fermionic contractions modify the weight
in front:
\begin{eqnarray}
{\cal O} (z_1, \bar{z}_1) {\cal O}^{-1} (z_2, \bar{z}_2) \sim
\frac{1}{|z_{12}|^{1/2}} \left[ \frac{4/3}{z_{12}} A(z_2) +
\frac{4/3}{\bar{z}_{12}} \bar{A}(\bar{z}_2) + 4 {\cal I} + 2 \tilde{D}   
(z_2, \bar{z}_2) + \ln |z_{12}| \tilde{C} (z_2, \bar{z}_2)  \right]
\label{opeOO-1}
\end{eqnarray}

This gives the new scaling formula
\begin{eqnarray}
\langle [{\cal O}{\cal O}^{-1}](1) [{\cal O}{\cal O}^{-1}](2) \rangle
\propto z^2 + \bar{z}^2 
\end{eqnarray}
to be contrasted with what was obtained in \cite{MudryNPB466}.  The other
scalings mentioned there will also be modified for $N=2$.  The correct
expressions are given below.

 The OPE of the local DOS are extracted from Eq.(\ref{raw}) (we
normalize the ${\cal M}$ two-point function).  First, we can
write the behaviour of (\ref{raw}) as $z_1 - z_2 = \epsilon
\rightarrow 0$:
\begin{eqnarray}
\langle [{\cal M M}](2) {\cal M}(3) {\cal M}(4) \rangle =
\frac{1}{|\epsilon z_{34}|^{1/2}} \left[ 1 + \gamma
\ln{|\epsilon|} + \gamma \ln{\left| \frac{z_{34}}{z_{23} z_{24}}
\right|} + ... \right]
\end{eqnarray}
in which $\gamma$ is a nonzero constant whose exact value is irrelevant for
our purposes (it could be obtained by carefully expanding
(\ref{raw})).  
The crucial fact is that
the operator in the symmetric representation, with weights $(-1, -1)$,
does not appear here again, like the adjoint operator in
(\ref{opehh-1}).  There is one 
quick way of seeing this last result.  As $z \rightarrow 0$, $\frac{z}{z-1}
\sim -z$.  Thus, summing the conformal blocks for $z$ and
$\frac{z}{z-1}$ amounts to adding the OPE (\ref{opeOO-1}) for $z$ and
$-z$, which makes the antisymmetric $1/z$ part disappear.  The consequence 
of this, when one considers the point-splitting procedure leading to
${\cal M}^{q}$, is that negative- dimensional operators simply do not
come about in the multifractality for $SU(2)$.

Postulating for ${\cal M}$ an OPE of the form (we remove the tildes
from the logarithmic operators, since we use a new normalization)
\begin{eqnarray}
{\cal M} (1) {\cal M}(2) \sim \frac{1}{|z_{12}|^{1/2}} \left[
 {\cal I} + D(2) + \frac{1}{2} \ln |z_{12}| C(2) + ... \right]
\label{opeMM}
\end{eqnarray}
then yields
\begin{eqnarray}
\langle C(1) {\cal M}(2) {\cal M}(3) \rangle &=& 2\gamma
\frac{1}{|z_{23}|^{1/2}} \nonumber \\
\langle D(1) {\cal M}(2) {\cal M}(3) \rangle &=& \gamma
\frac{1}{|z_{23}|^{1/2}} \ln{\left| \frac{z_{23}}{z_{12}z_{13}}
\right|} \nonumber \\ \nonumber \\
\langle D(1) D(2) \rangle &=& -2\gamma \ln{|z_{12}|} \nonumber \\
\langle D(1) C(2) \rangle &=& 2 \gamma \nonumber \\
\langle C(1) C(2) \rangle &=& 0
\end{eqnarray}
in turn leading to the OPEs
\begin{eqnarray}
D(1) D(2) = -2 \gamma \ln{|z_{12}|} + ..., ~~~ D(1) C(2) = 2 \gamma +
...,~~~ C(1) C(2) = 0 \nonumber \\ 
D(1) {\cal M} (2) =  -\gamma \ln|z_{12}| {\cal M} (2) + ..., ~~~
C(1) {\cal M} (2) = 2 \gamma {\cal M} (2) + ... \label{opeCDM}
\end{eqnarray}
which will be used later to infer the behaviour of the
multifractality. 
 
\section{$N > 2$ case}

Let us now outline what happens for the $N > 2$ case.  
We have seen in the previous section that, for $SU(2)$, the spectrum
of operator dimensions is cut right at
the beginning by the logarithmic locality conditions.  We can wonder
if the same holds in the $N>2$ case. 

The conformal
blocks, solutions to the $SU(N)_{-2N}$ Knizhnik-Zamolodchikov
equations, are ($F_i^p (z) = [z(1-z)]^{1-1/N^2} \tilde{F}_i^p (z)$)
\begin{eqnarray}
&\tilde{F}_1^a (z) &= \sqrt{2}N (1-z) F(2-1/N, 2+1/N; 2; z) \nonumber \\
&\tilde{F}_1^b (z) &= (1-z) F(2-1/N, 2+1/N; 3; 1-z) \nonumber \\
&\tilde{F}_2^a (z) &= z F(2-1/N, 2+1/N; 3; z) \nonumber \\
&\tilde{F}_2^b (z) &= \sqrt{2}N z F(2-1/N, 2+1/N; 2; 1-z)
\end{eqnarray}
For the case $N=2$, we had that the analytic continuations to $1/z$
were of logarithmic type.  Here, however, the correct single-valued
construction is like the standard one \cite{KnizhnikNPB247}, since the
analytic continuations 
\begin{eqnarray}
&\tilde{F}_1^a (1/z) &= - \sqrt{2}N z^{-1} (1-z) F(2-1/N, 2+1/N; 2; 1/z)
\nonumber \\
&\tilde{F}_1^b (1/z) &= - z^{1+1/N} (1-z) F(1+1/N, 2+1/N; 3; z)
\nonumber \\
&\tilde{F}_2^a (1/z) &=  z^{-1} F(2-1/N, 2+1/N; 3; 1/z) \nonumber
\\
&\tilde{F}_2^b(1/z) &= \sqrt{2}N z^{1+1/N} F(1/N, 2+1/N; 2; 1-z)  
\end{eqnarray}
are free of logarithms near $z = 0$ for $N \neq 2$.  On the other
hand, we have for $\frac{z}{z-1}$ the blocks
\begin{eqnarray} 
&\tilde{F}_1^a (\frac{z}{z-1}) &= \sqrt{2}N (1-z)^{1+1/N} F(1/N, 2+
1/N; 2; z) \nonumber \\ 
&\tilde{F}_1^b (\frac{z}{z-1}) &= (1-z)^{-1} F(2-1/N, 2+1/N; 3;
\frac{1}{1-z}) \nonumber \\  
&\tilde{F}_2^a (\frac{z}{z-1}) &= -(1-z)^{1+1/N} z F(1+1/N, 2+ 1/N; 3; z)
\nonumber \\ 
&\tilde{F}_2^b (\frac{z}{z-1}) &= - \sqrt{2}N (1-z)^{-1} z
F(2-1/N, 2+1/N; 2; \frac{1}{1-z}). \nonumber \\
\end{eqnarray}
The full correlator reads in this case
\begin{eqnarray}
\langle {\cal M}(1) {\cal M}(2) {\cal M}(3) {\cal M}(4) \rangle \sim
\frac{|1-z|^{2-2/N^2}}{|z_{12}z_{34}|^{2/N^2}} \left[ {\cal H}_{11} +
\frac{\bar{z}}{1-\bar{z}} {\cal H}_{12} + \frac{z}{1-z} {\cal H}_{21}
+ \frac{|z|^2}{|1-z|^2} {\cal H}_{22} \right]
\label{rawN}
\end{eqnarray}
where
\begin{eqnarray}
{\cal H}_{11} (z, \bar{z}) &&= H^{\prime}_{11}(z, \bar{z}) + |z|^{-2+4/N^2}
H^{\prime}_{11} (\frac{1}{z}, \frac{1}{\bar{z}}) + |1-z|^{-4(1-1/N^2)}
H^{\prime}_{11} (\frac{z}{z-1}, 
\frac{\bar{z}}{\bar{z}-1}) \nonumber \\  
{\cal H}_{12} (z, \bar{z}) &&= H^{\prime}_{12}(z, \bar{z}) - z
|z|^{-4(1-1/N^2)}
H^{\prime}_{12} (\frac{1}{z}, \frac{1}{\bar{z}}) - (1-\bar{z})
|1-z|^{-4(1-1/N^2)} H^{\prime}_{12} (\frac{z}{z-1}, 
\frac{\bar{z}}{\bar{z}-1}) \nonumber \\  
{\cal H}_{21} (z, \bar{z}) &&= H^{\prime}_{21}(z, \bar{z}) - \bar{z}
|z|^{-4(1-1/N^2)} H^{\prime}_{21}  (\frac{1}{z}, \frac{1}{\bar{z}})
- (1-z) |1-z|^{-4(1-1/N^2)} H^{\prime}_{21} (\frac{z}{z-1},
\frac{\bar{z}}{\bar{z}-1}) \nonumber \\  
{\cal H}_{22} (z, \bar{z}) &&= H^{\prime}_{22}(z, \bar{z}) + |z|^{-4(1-1/N^2)}
H^{\prime}_{22} (\frac{1}{z}, \frac{1}{\bar{z}}) + |1-z|^{-2+4/N^2}
H^{\prime}_{22} (\frac{z}{z-1}, 
\frac{\bar{z}}{\bar{z}-1}) 
\end{eqnarray}
with the correlators $H^{\prime}_{ij}$ given by
\begin{eqnarray} 
H^{\prime}_{ij}(z, \bar{z}) =   \tilde{F}_i^a (z)
\tilde{F}_j^b(\bar{z}) + \tilde{F}_i^b (z) \tilde{F}_j^a
(\bar{z}).
\label{completefourpointfunctionN}
\end{eqnarray}
As for the $SU(2)_{-4}$ case, the $b$ contour
solutions have logarithmic behaviour near $z=0$ for $z$  and
$\frac{z}{z-1}$ (see appendix).  Moreover, this logarithmic behaviour
is exactly of the same form, involving the same powers of $z$.  Thus,
again, the operator in the adjoint representation, having conformal
weights $(-1, -1)$, does not appear in the OPE of the decoupling field
with its inverse, because of single valuedness.

However, when one considers the OPE of $h$ with $h$, for which the
analytic continuation to $1/z$ is necessary, the difference with the
$SU(2)_{-4}$ becomes clear.  In fact, the analytic continuations of
$\tilde{F}$ to $1/z$, in contrast to the $N=2$ case, are completely
regular for $N > 2$, as we have seen above.  Expanding the full
correlator (\ref{rawN}) around $z=0$, we find that the operators in
the symmetric (S) and antisymmetric (A) representations {\it do} appear, as
proposed in \cite{MudryNPB466}.  These have conformal weights
\begin{eqnarray}
h_S = \bar{h}_S = -\frac{(N+2)(N-1)}{N^2}, \nonumber \\
h_A = \bar{h}_A = -\frac{(N-2)(N+1)}{N^2}
\end{eqnarray}
Thus, we now have {\it a negative dimensional operator on the physical
level}, in contrast to the $N=2$ case.  This operator has conformal
weights
\begin{eqnarray}
h = \bar{h} = \frac{-1}{N}(1-2/N)
\end{eqnarray}
For $N=2$, this vanishes, in accordance with the work of the previous
sections.  

Logarithms in conformal field theories originate from degeneracies in
the spectrum of conformal dimensions.  When two or more operators have
weights which become degenerate, they become distinct again by
incorporating powers of $\ln z$, which is a weight zero object.  For
the $N=2$ case, three operators in the theory were of weight zero: the
unit operator, the first descendant of the adjoint operator, and the
operator in the antisymmetric representation.  This high level of
degeneracy has produced the logarithmic degeneracies which, through
the requirements of single-valuedness (locality) of the correlator,
have led us to discover the termination of the spectrum.  

In the case of $N>2$, we have just seen that the termination does {\it
not} appear immediately.  We can however speculate about the moment at
which the termination should occur.  This 
relies on basic arguments of representation theory.  For
$SU(N)_{-2N}$, the conformal weights of the operators in a given
representation characterised by a Young tableau with $f_i$ boxes in the
$i$-th row is given by the associated Casimir
\begin{eqnarray}
c_{\{f_i\}} = \frac{1}{2} \sum_{i=1}^N [f_i^2 +(N+1-2i)f_i] -
\frac{f^2}{2N}, ~~~~~ f = \sum_{i=1}^N f_i
\end{eqnarray}
as (considering holomorphic and anti-holomorphic)
\begin{eqnarray}
h_{\{f_i\}} = -\frac{c_{\{f_i\}}}{N}, ~~~
\bar{h}_{\{\bar{f}_i\}} = -\frac{\bar{c}_{\{\bar{f}_i\}}}{N}
\end{eqnarray}
Now it is easy to check that for a completely antisymmetric
representation of $SU(N)$, i.e. $f_i =1, i=1,...,N$, the value of the
Casimir vanishes.  The associated operator thus has vanishing conformal
weights, which makes it degenerate with the unit operator and the
descendant of the adjoint operator, like in the $N=2$ case extensively
presented above.  This operator in the completely antisymmetric
representation is generated only by a product of $N$ fundamental
representations (our basic operator), so will appear only for $q = N$.
We thus speculate that, for $N>2$, the additional conformal weight
degeneracy at $q=N$ leads to logarithmic degeneracies which, exactly in
the same way as previously, lead to the termination of further scaling.

\section{Termination of the parabolic multifractal spectrum}

The relationship between the operator dimensions in the CFT and the
multifractal spectrum were derived with the help of real-space
renormalization group arguments in \cite{MudryNPB466}, where all the
definitions and notations used here are put forward.  Three types of
multifractal spectra are defined: $\tau(q)$, which is the disorder
averaged logarithm of the inverse participation ratio of normalized
wavefunctions;  $\tau^*(q)$ (the one we calculate here), which is the
logarithm of the ratio of disorder averaged unnormalized wavefunctions
and disorder averaged normalization to the appropriate power, and
$\tau^{**}(q)$, which is the logarithm of the disorder averaged
participation ratio of normalized wavefunctions.  There exist strong
constraints on the first of these, $\tau(q)$.  It can only be a
monotonously increasing function of $q$, which forces the cut of the
parabolic approximation at a given point.  The other functions have
not been shown to have to obey to such conditions.

The only multifractal spectrum that can be calculated with the help of
the present theory is $\tau^*(q)$.  It is related to the conformal
weights $(h_q, \bar{h}_q)$ of the most relevant operator contained in
the point-splitting definition of ${\cal M}^q$ according to 
\begin{eqnarray}
\tau^*(q) = (q-1) (2-\Delta_1) + \Delta_q - \Delta_1
\end{eqnarray}
where $\Delta_q = h_q + \bar{h}_q$.  These weights are obtained by
considering the decay of the correlator
\begin{eqnarray}
\langle {\cal M}^q (1) {\cal M}^q (2) \rangle \sim z_{12}^{-2h_q}
\bar{z}_{12}^{-2\bar{h}_q} 
\end{eqnarray}
Let us first recall the arguments used by \cite{MudryNPB466} to obtain
these weights.  The operator ${\cal O}$ (\ref{localoperators}) is made
from the product of free fermions and of a field in the fundamental
representation of $SU(N)_{-2N}$.  The fusion of the fermions is
straightforward, but the fusion of the WZNW primary fields is a bit
less obvious.  Building the Young tableaux associated to the tensor
product of $q$ fundamental representations and evaluating the
quadratic Casimir eigenvalues $c_m$ of all the resulting irreducible
representations, one obtains a list of operators having conformal
weights \cite{KnizhnikNPB247,MudryNPB466} $h_m = -c_m/N$.  It is then
easy to identify the {\it most relevant} operator as the one in the
completely symmetric representation, which has $h_S = -\frac{N-1}{2N^2}
q(q+N)$.  The resulting combined fermionic and WZNW weights lead to
the provisional result \cite{MudryNPB466} 
\begin{eqnarray}
h_q = \bar{h}_q = \frac{q}{2} - \frac{N-1}{2N^2}(q^2 + Nq) ~~~~~ (provisional)
\label{hmudry} 
\end{eqnarray}
and led to the corresponding scaling exponents  $\tau^*(q)$ given by
\begin{eqnarray}
\tau^*(q) = (q-1) \left(2-\frac{N-1}{N^2}q\right) ~~~~~ (provisional)
\label{taumudry}
\end{eqnarray}
One might question the fact that this $\tau^*(q)$ becomes arbitrarily
negative for large values of $q$.  It attains its maximum value at
$q_{max} = N(1-1/N)^{-1} + 1/2 > N$, after which the parabola bends
back down.  Even though it is in principle not impossible that
$\tau^*(q)$ could take on negative values (see the discussion in
\cite{MudryNPB466}), a limit to this behaviour would be most welcome.  

However, we can see from the previous sections that the
straightforward WZNW fusion rules are not applicable in our case.
Namely, we can perform for example the explicit point-splitting
procedure leading to ${\cal M}^3$ for the case $N=2$ from the fusion
rules (\ref{opeCDM}):
\begin{eqnarray}
{\cal M}^3 (z) = \lim_{a \rightarrow 0} {\cal M}(z+a) {\cal M}(z)
{\cal M}(z-a) = \lim_{a \rightarrow 0} |a|^{-1/2} \left[ {\cal I} +
D(z) + \frac{1}{2} \ln{|a|} C(z) + ... \right] {\cal M}(z-a) \nonumber
\\ = (\lim_{a
\rightarrow 0} |a|^{-1/2}) {\cal M}(x) + ... \sim {\cal M}(x)
\end{eqnarray}
showing explicitly that $\langle {\cal M}^3 (1) {\cal M}^3 (2)
\rangle \sim |z_{12}|^{-1/2}$, i.e. that $h_3 = \bar{h}_3 = 1/8$, in
contrast to the results of \cite{MudryNPB466}, for which $h_3 =
\bar{h}_3 = -3/8$.  Generalizing to arbitrary $q$ (still for the
specific case of $N=2$), we obtain that 
\begin{eqnarray}
{\cal M}^{2p} \propto {\cal I}, ~~~~~ {\cal M}^{2p +1} \propto {\cal
M}
\end{eqnarray}
as far as the correlators $\langle {\cal M}^q (1) {\cal M}^q (2)
\rangle$ are concerned.  This implies the explicit termination of the
parabolic law for the multifractality in the case of $SU(2)$ vector
potential randomness, which can be understood as a vanishing of OPE
coefficients associated to operators with more relevant dimensions.
This type of termination seems special to the $SU(N)$ non-Abelian
disorder case, since the Abelian multifractality, when computed from
CFT, only involves expectation values of vertex operators of Gaussian
distribution which do not carry the same conformal block structure as
the ones that we have here.  Note also that in the case of the
$gl(1,1)$ WZNW model \cite{RozanskyNPB376}, fields with arbitrarily
negative conformal dimensions can be obtained from the fusion of a
sufficient number of fundamental representations.  These interesting
differences still need to be more properly understood. 

Finally, from the results and discussion of the previous
section, we can conjecture that the multifractal spectrum
(\ref{taumudry}) for general $N$ should be complemented by a condition
of the sort  
\begin{eqnarray}
q < q_{c} 
\end{eqnarray}
where $q_c \lesssim N$ is the order at which the proper treatment of
the locality conditions in the presence of logarithmic degeneracies
kill off further scaling.  Thus, above this critical value
of $q$, the parabolic law will not hold anymore.  From our explicit
results for $SU(2)$, and comparing with the results from
\cite{MudryNPB466}, we can say that $2 < q_c (SU(2)) < 3$.  This is
consistent with the exact solution for $\tau(q)$ itself \cite{CauxTBP}.

\section{What is wrong with replicas ?}

In general, the theoretical treatment of disordered systems is a
notoriously difficult task in view of the fact that it is necessary
to average the {\it logarithm} of the partition function over the
statistical ensemble in order to obtain physical results.  Two general
approaches have been developed to perform such types of calculations.
First and foremost, the supersymmetry (SUSY) approach \cite{Efetov97}
is a mathematically sound technique that makes use of commuting and
anticommuting fields simultaneously, making the partition function $Z$
equal to one by definition.  Its only setback is its limited
applicability: interacting disordered systems cannot be represented by
a SUSY field theory.

The alternative technique is the well-known replica approach,
introduced in \cite{EdwardsJPF5}, which allows to rewrite the average of
$\ln Z$ as a tractable object by making use of the identity
\begin{eqnarray}
\overline{\ln Z} = \lim_{r \rightarrow 0} \frac{\overline{Z^r} -1}{r}
\end{eqnarray}
i.e. by considering a replicated theory with $r$ species of the
original field.  One then hopes that the physical properties of the
disordered system are faithfully represented by the analytical
continuation to $r \rightarrow 0$.  However, the main problem is that
the replica approach is not mathematically well-defined.  Indeed, even
though it has been used successfully quite often, the pathologies
persist (for example, the systems obtained by
choosing a positive or negative number of replicas are markedly
different, making the limit $r \rightarrow 0$ completely ill-defined).
It has been argued in \cite{VerbaarschotJPA17} that failure of the
replica trick may occur whenever the theory for a general integer
value of $r$ possesses different symmetries than the one at $r=0$.

We are thus now in a position to compare our present exact results
with the previous ones obtained
with the replica approach \cite{CauxNPB466}.  The conformal blocks obtained 
there were solutions to $SU(0)_N$ equations, and different to the 
ones obtained here.  The OPE of the physical field ${\cal M}$ ($Q$ in
\cite{CauxNPB466}) was given by
\begin{eqnarray}
{\cal M} (1) {\cal M} (2) \sim \frac{1}{|z_{12}|^{2/N^2}} \left[ {\cal I} -
z (D(2) + C(2) \ln |z|) - \bar{z} (\bar{D}(2) + \bar{C}(2) \ln |z| ) +
... \right]
\end{eqnarray}
where the chiral-like logarithmic operators had correlators given by
the expressions
\begin{eqnarray}
&&\langle D(1) D(2) \rangle \sim \frac{1}{z_{12}^2} [\ln |z_{12}| + c]
\nonumber \\
&&\langle D(1) C(2) \rangle \sim \frac{1}{z_{12}^2} \nonumber \\
&&\langle C(1) C(2) \rangle  = 0
\end{eqnarray}
with similar expressions for $\bar{C}, \bar{D}$ given by $z$
substituted by $\bar{z}$.  Although the basic two-point function
generated from this OPE will be the same as the one obtained through
above, the higher-point functions will show marked differences with
the exact ones, which makes all further arguments unreliable. 

Thus, although the replica approach successfully accounts for the basic
dimensionality of the operators, it does not correctly describe their
higher-level correlations.  The failure of the replica solution may be
understood in terms of the interpretation put forward in
\cite{VerbaarschotJPA17}, since the essential ingredient in the replica
solution was the $SU(r)$ symmetry (associated to the $r$ different
flavours of replicated fermions), which obviously is a different
symmetry at $r=0$ than at integer $r$.

\section{Conclusion}

Let us give a brief summary of our results. 

(a) We have demonstrated explicitly that for the $SU(2)$ group, the
basic WZNW fusion rules are modified by the proper treatment of the
locality conditions in the presence of so-called logarithmic
operators, such that operators with negative conformal dimensions do
not appear in OPE of the local density of states that we are
interested in for multifractality purposes;

(b) We have provided arguments based on the nature of the conformal
field theory and its operator content, to conjecture that the
mechanism for the termination of the multifractal spectrum that we
propose should have occurred as well before $q=N$ at the latest for
general $SU(N)$.  In these theories, there will thus be a finite
string of physical operators with negative dimensions influencing the
multifractality, but not an infinite set as previously believed.  The
termination for $N>2$ is conjectured to occur by the above mechanism
before the scaling exponents $\tau^*(q)$ reach their maximum value,
which provides evidence that the multifractal spectrum $\tau^*(q)$
obeys the same type of inequalities as the spectrum $\tau(q)$ defined
with normalized wavefunctions;

(c)  We have shown that the replica approach fails to reproduce the
correct nonperturbative multipoint correlation functions in the
conformal limit.

\section{Acknowledgments}

We are grateful to I. Lerner, K. Efetov, V. Fal'ko, V. Kravtsov, C.
Mudry and especially I. Kogan for many interesting discussions,
constructive criticism and interest to the work. We also should like to
thank the ITP, Santa Barbara, where the early discussions leading to
this work took place under the support of NSF grants DMR-92-23217 and
PHYS94-07194.  J.-S. Caux acknowledges support from NSERC Canada, and
 from the Rhodes Trust.  N. Taniguchi acknowledges support from the Japan
Society for the Promotion of Science.

\section{Appendix 1: the Haar Measure}
For the sake of clarity, we here include the derivation of the Haar
measure on $SU^C(2)/SU(2)$ as can be found in \cite{Helgason84}.

For a function $f(g^C) \in C_c(G^C)$, where $C_c(G^C)$ is the space of
continuous functions of compact support on $G^C$, there exists a
function $D(k,a,n)$ such that the group
integration, after the Iwasawa decomposition, can be written as
\begin{eqnarray}
\int_{G^C} f(g^C) d g^C = \int_{KAN} f(kan) D(k,a,n) dk da dn
\end{eqnarray}
Since $G^C$ is unimodular (i.e. its left invariant measure is also
right invariant), the LHS does not change when we replace $f(g^C)$ by
$f(k_1an_1), k_1 \in K, n_1 \in N$.  Then, it follows that
$D(k_1^{-1} k, a, n n_1^{-1}) \equiv D(k,a,n)$, so $D(k,a,n)$ is a
function $\delta(a)$ of $a$ only.  Therefore, for $a_1 \in A$, we have
\begin{eqnarray}
&& \int_{G^C} f(g^C) dg^C = \int_{G^C} f(g^Ca_1) dg = \int_{KAN}
f(kana_1) \delta(a) dk da dn \nonumber \\ 
&& \quad = \int_{KAN} f(kaa_1(a_1^{-1}na_1)) \delta(a) dk da dn =
\int_{KAN} f(ka(a_1^{-1} na_1)) \delta(aa_1^{-1}) dk da dn 
\nonumber \\ 
&& \quad = \int_{KAN} f(kan) \delta(aa_1^{-1}) dk da I(a_1)^* (dn)
\end{eqnarray}
where $I(a_1)$ is the automorphism $n \to a_1 n a_1^{-1}$ of $N$, and
$I^*\omega$ denotes the transform (pullback) of $\omega$ by $I$.  We
have made use in the last step of the equality
\begin{eqnarray}
\int_{N} (f \circ  \Phi^{-1})\omega = \int_{M} f \Phi^* \omega
\end{eqnarray}
in which $M$ and $N$ are two oriented manifolds, and $\Phi$ is a
diffeomorphism between them.

Finally, by looking at
\begin{eqnarray}
a_1 n a_1^{-1} = \left(\begin{array}{cc} 
1 & [\mu_1 + i\mu_2]e^{\phi_1} \\ 0 & 1 \end{array}\right)
\end{eqnarray}
we see that $I(a_1)^*(dn) = e^{2\phi_1} d\mu_1 d\mu_2 = e^{2\phi_1}
d\mu_+ d\mu_-$, and choosing $a_1 = a$ produces the Haar measure
(\ref{haarmeasure}).

\section{Appendix 2:  Analytic Continuations}

We here give for reference the analytic continuation formulas used in
the text, as well as the expansions of the relevant functions around
$z=0$.

The second solutions for the conformal blocks, labeled by their
contour $b$, are hypergeometric functions of the argument $1-z$.  When
$z \to 0$, these behave logarithmically.  The proper expressions
become
\begin{eqnarray}
&& \tilde{F}_1^b (z) = \frac{4}{\pi} \left[\frac{4/3}{z} +
\frac{\tilde{F}_1^a (z)}{2\sqrt{2}} \ln z - 4/3 + (1-z) K_{11}(z)
\right] \nonumber \\ 
&& \tilde{F}_2^b (z) = \frac{\sqrt{2}}{\pi} \left[\frac{16/3}{z} +
\tilde{F}_2^a (z) \ln z + 4/3 + z K_{12} (z) \right] \nonumber \\
&& F(1/2, 5/2; 2; 1-z) = \frac{-1}{\pi} \left[ \frac{-4/3}{z} + F(1/2,
5/2; 2; z) \ln z + K_{01} (z) \right]  
\end{eqnarray}
where
\begin{eqnarray}
K_{ij} (z) &&= \sum_{n=0}^{\infty} z^n \frac{(1/2+i)_n (5/2)_n}{(n+j)!
n!} \left[\psi(n+i+1/2) + \psi(n+5/2) - \psi(n+j+1) - \psi(n+1)\right] 
\end{eqnarray}

Near $z=0$, we can then perform the expansions
\begin{eqnarray}
\tilde{F}_1^a (z) &=& 2\sqrt{2} [1 + 7/8 z + O(z^2)] \nonumber \\
\tilde{F}_1^b (z) &=& \frac{4}{\pi} \left[ \frac{4/3}{z} + \ln z -4/3
+ K_{11}(0) \right] + O(z\ln z) \nonumber \\
\tilde{F}_2^a (z) &=& z + 5/4 z^2 + O(z^3) \nonumber \\ 
\tilde{F}_2^b (z) &=& \frac{\sqrt{2}}{\pi} \left[ \frac{16/3}{z} - 4/3
\right] + O(z \ln z) \nonumber 
\end{eqnarray}
\begin{eqnarray}
&& F(1/2, 5/2; 2; z) = 1 + 5/8 z + O(z^2) \nonumber \\
&&  F(1/2, 5/2; 2; 1-z) = \frac{-1}{\pi} \left[ \frac{-4/3}{z} + \ln z
+ K_{01}(0) \right]  + O(z \ln z) 
\end{eqnarray}
used in the determination of the OPEs in the text.

For the $N >2$ case, we have for example
\begin{eqnarray}
\tilde{F}_1^b (z) = \frac{2}{\Gamma(1-1/N) \Gamma(1+1/N)} \left[
\frac{N^2}{(N^2-1)z} + \tilde{F}_1^a (z) \ln z - \frac{N^2}{N^2-1} +
(1-z) J (z) \right] 
\end{eqnarray}
with $J(z)$ some function regular as $z \to 0$.

\end{document}